\documentclass[aps,preprintnumbers,amsmath,amssymb,superscriptaddress,floatfix,nofootinbib]{revtex4}
\usepackage{lipsum}
\usepackage{cancel}
\usepackage{slashed}
\usepackage{graphicx}
\usepackage{epsfig}
\usepackage{epstopdf}
\usepackage{hyperref}
\usepackage{amsmath}
\usepackage{amsfonts}
\usepackage{amssymb}
\usepackage{setspace}
\usepackage{amsfonts,color}
\usepackage{caption}
\usepackage{natbib}
\usepackage{subfig}
\usepackage{color}
\usepackage{bbold}
\usepackage{bm}
\usepackage{mathrsfs}
\usepackage{url}
\usepackage{rotating}
\usepackage{pdflscape}
\usepackage{diagbox}
\usepackage{soul}
\usepackage[normalem]{ulem}

\usepackage{xcolor}
\usepackage[force]{feynmp-auto}

\usepackage{multirow}
\usepackage{hhline}

\begin{document}

\title{Effects of odderon spin on helicity amplitudes in $pp$ elastic scattering}

\author{Prin Sawasdipol}
\email{p.namwongsa@kkumail.com}
\affiliation{Khon Kaen Particle Physics and Cosmology Theory Group (KKPaCT), Department of Physics, Faculty of Science, Khon Kaen University,123 Mitraphap Rd., Khon Kaen, 40002, Thailand}
\author{Jingle B. Magallanes}
\email{jingle.magallanes@g.msuiit.edu.ph}
\affiliation{Department of Physics, Mindanao State University - Iligan Institute of Technology, Iligan City, 9200, Philippines}
\author{Chakrit Pongkitivanichkul}
\email{chakpo@kku.ac.th}
\affiliation{Khon Kaen Particle Physics and Cosmology Theory Group (KKPaCT), Department of Physics, Faculty of Science, Khon Kaen University,123 Mitraphap Rd., Khon Kaen, 40002, Thailand}
\author{Daris Samart}
\email{darisa@kku.ac.th: corresponding author}
\affiliation{Khon Kaen Particle Physics and Cosmology Theory Group (KKPaCT), Department of Physics, Faculty of Science, Khon Kaen University,123 Mitraphap Rd., Khon Kaen, 40002, Thailand}

\date{\today}

\begin{abstract}

In recent years, the discovery of the odderon, a colorless $C$-odd gluonic compound, has been confirmed in the TOTEM and D0 collaborations. However, the spin quantum number of the odderon remains unidentified. In this work, we aim to attribute a spin of $J=3$ to the odderon in $pp$ elastic scattering by calculating the helicity amplitudes and the corresponding complex parameter $r_5$, the ratio of helicity's single-flip to non-flip amplitudes, for the spin-3 tensor odderon with the standard spin-2 tensor pomeron exchanges. Then, we apply these results to the constraints obtained from the STAR experiment at RHIC. By comparing to the contributions of the spin-1 vector odderon and spin-2 tensor pomeron, we demonstrate that the spin-3 tensor odderon, i.e. $J=3$, provides a better explanation for the observable in $pp$ elastic scattering.

\end{abstract}


\maketitle


\newpage
\section{\label{sec:introduction}Introduction}


In soft high-energy collisions, specifically hadron-hadron elastic scattering with a large center-of-mass energy, $\sqrt{s}$, and small momentum transfer, $\sqrt{|t|}$, the Regge theory of strong interaction phenomenology has been widely used to explain these processes. The exchange of particles mediated by the pomeron trajectory, as predicted by Regge theory, has proven to be successful in describing such interactions \cite{Regge1959:14-951}. The pomeron, which corresponds to vacuum quantum number states of the reggeon, accurately accounts for the rising behavior observed in the total cross-section of high-energy hadronic collisions \cite{Collins1977-intro-regge, Collins1984-hadron, Gribov2008-strong-interaction}. Additionally, it satisfies the Froissart bound \cite{Froissart:1961ux, Lukaszuk1967:52-122}, which imposes an upper limit on the total cross-section of two-particle scattering at high energy. The Froissart bound arises from the Regge trajectory's vacuum quantum numbers. Furthermore, the Regge trajectories often yield a Chew–Frautschi plot \cite{Chew1962:8-41}, which reveals the relationship between mass and spin. This significant plot is utilized to effectively identify higher spin Regge trajectories.

Nevertheless, when considering $pp$ and $p\bar{p}$ scattering processes, a model based solely on pomeron exchange is found to be inconsistent with the available data. The emergence of Quantum Chromodynamics (QCD) provided theorists with a framework to propose the inclusion of odderon exchange alongside pomeron exchange, aiming to explain the observed rising behavior in the total cross section for $pp$ scattering \cite{Donnachie2002-pomeron, Barone2002-diffraction}.
The odderon was initially predicted in 1973 as a composite state of three reggeized gluons, based on the asymptotic theorem proposed by L. Lukaszuk and B. Nicolescu \cite{Lukaszuk:1973nt}. In detail, the odderon contributes to scattering amplitude by decomposing the amplitude into parts associated with charge-conjugation exchanges, specifically $C=+1$ for the pomeron and $C=-1$ for the odderon. Two decades later, the theory of the odderon was further developed within perturbative QCD, describing it as a bound state of three gluons \cite{Bartels:1999yt}. Subsequently, the pomeron and odderon have been studied through various approaches, including phenomenological Regge theory 
\cite{Ewerz:2013kda,Ewerz:2016onn, Covolan:1996uy,Block:2012nj,Szanyi:2019kkn,Jenkovszky:2020jca,Broniowski:2018xbg,Csorgo:2020wmw,Csorgo:2019ewn,Csorgo:2018uyp,Ster:2015esa,Block:1984ru,Khoze:2018bus}, QCD-inspired models 
\cite{Halzen:1992vd,Donnachie:1983ff,Ma:2001ji,Hu:2002zr,He:2003fq,Hu:2008zze,Zhou:2006zx,Lu:2020wpo,Bartels:1999yt}, and string theory or AdS/CFT correspondence 
\cite{Domokos:2009hm,Domokos:2010ma,Avsar:2009hc,Hu:2017iix,Xie:2019soz,Burikham:2019zbo,Liu:2022zsa,Liu:2022out,Ballon-Bayona:2017vlm,Iatrakis:2016rvj}.

{ In 2021, the TOTEM and D0 collaborations made a significant discovery regarding the existence of odderon exchange in $t$-channel $pp$ and $p\bar{p}$ scattering at high energies, with a combined significance exceeding 5$\sigma$ \cite{TOTEM:2020zzr}. The analysis, based on a model-independent approach, compared the $p\bar{p}$ elastic cross section at $\sqrt{s} = $ 1.96 TeV, as measured by the D0 collaboration, with the extrapolated values obtained from $pp$ scattering at $\sqrt{s} = $ 2.76, 7, 8, and 13 TeV, as measured by the TOTEM collaboration. This comparison confirmed the presence of odderon exchange. Subsequently, a theoretical study, employing model-dependent methods, further increased the statistical significance to at least 7.08$\sigma$ by comparing the differential cross sections of elastic $pp$ scattering with the extrapolated $p\bar{p}$ scattering at LHC energies, specifically $\sqrt{s} = $ 2.76 and 7 TeV \cite{Csorgo:2020wmw}. }

Furthermore, it is crucial to investigate the spin structure of the pomeron and odderon. This is because, in elastic scattering, the evaluation of the scattering cross section requires determining the spin-dependent helicity amplitudes of the model, using unitarity and optical theorem. Indeed, a sophisticated framework for understanding the spin structure involves examining the complex parameter $r_5$, namely the ratio of spin-flip to non-flip amplitudes \cite{Goldberger:1960,Buttimore:1998rj}. A decade ago, the STAR experiment at RHIC conducted a study on single-spin asymmetry measurements and derived a strong constraint on $r_5$ as the central value of $(0.0017 + i\,0.007)$ with its statistical and statistical+systematic uncertainties at $\sqrt{s} = 200$ GeV \cite{STAR:2012fiw}. 
Therefore, to assess the viability of pomeron and odderon models, any spin-dependent observable from these models that are compatible with the constraint on $r_5$ can be regarded as a candidate model for pomeron and odderon contributions.

Studies on the spin structure of the odderon date back two decades when the relevant spin asymmetry $A_{NN}$ were demonstrated \cite{Leader:1999ua}.
In \cite{Ewerz:2016onn}, the possibility of pomeron spin contributions on the 
$r_5$ parameter has been studied based on helicity amplitudes with the Donnachie-Landshoff (DL) pomeron ansatz. Using experimental data from the STAR at RHIC \cite{STAR:2012fiw}, they show that the spin-2 tensor pomeron gives the most consistent value. Given the recent discovery of the odderon, the possibility of odderon contribution to the $r_5$ parameter has been tantalizing and become an active field of research. For example, the study on odderon contribution based on diquark model and QCD-inspired model can be found in \cite{Szymanowski:2016mbq} and \cite{Hagiwara2020:80-427} respectively. Our recent approach on odderon as a Regge oddball spin-3 with the DL propagator gives a theoretical basis for the odderon calculation and satisfies several experimental results in $pp$ and $p\bar p$ elastic scattering \cite{Magallanes2022-odderon}. 
Therefore in this paper, we provide an attempt to attribute the spin of the odderon as $J=3$ on helicity amplitudes in $pp$ elastic scattering by studying the combined exchange of pomeron and odderon based on the aspect of spin-dependence reaction.

The objectives of this study are as follows: (1) to calculate the helicity amplitudes of the pomeron and odderon based on Effective Field Theory (EFT) using DL ansatz for the pomeron and odderon propagator forms, (2) to calculate the corresponding complex parameter $r_5$ and then apply to the constraints of the STAR experiment at RHIC, and (3) to evaluate the possibility of attributing $J=3$ to the odderon that contributed with pomeron to the $pp$ elastic scattering.



We organize this paper as follows: Section \ref{sec:method} provides comprehensive details of the theoretical frameworks for the \emph{combined model} of spin-2 tensor pomeron and spin-3 tensor odderon. This model serves as the main study of our research. In addition, the \emph{combined model} of spin-2 tensor pomeron and spin-1 tensor odderon is also described and is included to complement the evaluation of our aim. In Section \ref{sec:result}, we present the results and engage in detailed discussions regarding the helicity amplitudes and the evaluation of the observable $r_5$ for both combined models. We also compare our findings to previous studies and assess the implications of our results. We finally summarize our findings and draw conclusions in the Section \ref{sec:conclusion} highlighting the contributions and potential future directions of research in this field.

\section{Theoretical Framework \label{sec:method}}



The transition amplitude of a scattering process is of great importance as it generally relates Quantum Field Theory (QFT) to observable in experiment. However, obtaining unpolarized cross section for any process requires squaring the amplitude, which leads to an exponential increase in the number of terms and renders the calculation of complex processes infeasible.

In contrast, the helicity approach directly calculates the transition amplitude by fixing the basis for the polarization states of external particles. This framework incorporates a dispersion approach for crossing relations \cite{Goldberger:1957}, a partial wave decomposition of the scattering amplitudes \cite{Jacob:1959}, and a connection of the dispersion relation to partial wave amplitudes \cite{Mandelstam:1958}. Together, these components form a complete theory of nucleon-nucleon elastic scattering for low energy, which is the momentum transfer in this work, denoted as $\sqrt{|t|}$.

\begin{figure}[!htbp]
\includegraphics[scale=0.5]{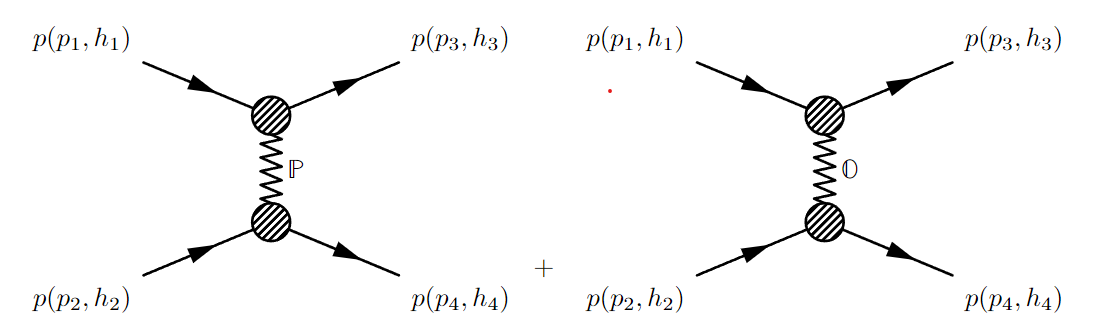}
\caption{\label{fig:fd-pp-scattering} Feynman diagram of $t$-channel $1+ 2 \rightarrow 3 +4$ elastic $pp$ scattering with the intermediate exchanges of pomeron $\mathbb{P}$ and odderon $\mathbb{O}$.}
\end{figure}

In this work, we considered the $pp$ elastic scattering process where $1, 2$ are incoming and $3, 4$ are outgoing particles. Each particle has its corresponding masses $m_i^{}$ four-momenta $p_i^{}$ and helicities $h_i$, where $i=1,..,4$ are the number of particles. According to our motivation in this work, moreover, the contribution of the pomeron and odderon exchanges are considered. The Feynman diagrams of the process are shown in the Fig. \ref{fig:fd-pp-scattering}.

The kinematics of the process are typically defined as
\begin{align}
\begin{aligned}
    s &= (p_1^{}+p_2^{})^2 = (p_3^{}+p_4^{})^2 = 4(\mathbf{p}^2 + m^2),  \\
    t &= (p_1^{}-p_3^{})^2 = (p_2^{}-p_4^{})^2 = -2\mathbf{p}^2(1 - \cos(\theta)),  \\
    u &= (p_1^{}-p_4^{})^2 = (p_2^{}-p_3^{})^2 = -2\mathbf{p}^2(1 + \cos(\theta)),
\end{aligned}
\end{align}
where $s+t+u = \sum_{i=1}^{4} m_i^2$ and $p=(E,\mathbf{p})$, where $E$ is its energy and $p$ its three-momentum. Additionally, for equal mass $m_i^{} = m$ and c.m. frame $\mathbf{p}_1^{} = \mathbf{p}_3^{} = \mathbf{p}$ where $-1 \le \cos\theta \le 1$ is the angle between the three-momenta of particles 1 and 3 in the c.m. frame, the physical region for $t$-channel scattering is then given by $t \ge 4m^2,~ u \le 0,~ s\le0$ \cite{Donnachie2002-pomeron}.

First, we constructed the \emph{combined spin-3 tensor-odderon} model by employing the contributions of the spin-2 tensor pomeron from \cite{Ewerz:2016onn} and spin-3 tensor odderon from \cite{Magallanes2022-odderon}. The Lagrangian given from those works read
\begin{align}
    \mathcal{L}_{\mathbb{P}_T^{}}^{}(x) &= - i\, g_{\mathbb{P}_T^{}}^{} \bar{\psi}_p^{}(x) \, \mathcal{G}_{(\mu\nu)}^{\mu'\nu'} \, \gamma_{\mu'} \, \overleftrightarrow{\partial_{\nu'}} \, \psi_p^{}(x) \, \mathbb{P}^{\mu\nu}(x), \label{eqn:pom-lag} \\
    \mathcal{L}_{\mathbb{O}_T^{}}^{}(x) &= - \frac{g_{\mathbb{O}_T^{}}^{}}{{M}_0^{}}\, \mathcal{G}_{(\mu\nu\rho)}^{\mu'\nu'\rho'} \, \bar{\psi}_p(x) \, \gamma_{\mu'}^{} \, \overleftrightarrow{\partial_{\nu'}^{}} \, \overleftrightarrow{\partial_{\rho'}^{}} \, \psi_p(x) \, \mathbb{O}^{\mu\nu\rho}(x), \label{eqn:odd-lag}
\end{align}
where $g_{\mathbb{P}_T^{}}^{}$ and $ g_{\mathbb{O}_T^{}}^{}$ are coupling constants and they carry the mass dimension as [mass]$^{-1}$, $\mathbb{P}^{\mu\nu}(x)$ and $\mathbb{O}^{\mu\nu\rho}(x)$ denote its descriptions, and $\mathbb{P}_T^{}$ and $\mathbb{O}_T^{}$ denote the spin-2 tensor pomeron and spin-3 tensor odderon, respectively. The mass parameter $M_0^{}$ is a free parameter and is introduced for correcting mass dimension. $\psi_p^{}$ is the Dirac field operator of the proton. $\overleftrightarrow{\partial_\mu^{}} \equiv \overleftarrow{\partial_\mu^{}} - \overrightarrow{\partial_\mu^{}}$. The totally symmetric operators given by
\begin{align}
    \mathcal{G}_{(\mu\nu)}^{\mu'\nu'} &= \frac{1}{2!} \left( g_{\mu}^{\mu'}g_{\nu}^{\nu'} + g_{\nu}^{\mu'} g_{\mu}^{\nu'} \right),  \\ 
    \mathcal{G}_{(\mu\nu\rho)}^{\mu'\nu'\rho'} &= \frac{1}{3!}\left( g_{\mu}^{\mu'} g_{\nu}^{\nu'} g_{\rho}^{\rho'} + g_{\nu}^{\mu'} g_{\rho}^{\nu'} g_{\mu}^{\rho'} + g_{\rho}^{\mu'} g_{\mu}^{\nu'} g_{\nu}^{\rho'} + g_{\nu}^{\mu'} g_{\mu}^{\nu'} g_{\rho}^{\rho'} + g_{\mu}^{\mu'} g_{\rho}^{\nu'} g_{\nu}^{\rho'} + g_{\rho}^{\mu'} g_{\nu}^{\nu'} g_{\mu}^{\rho'} \right),
\end{align}
are used to ensure the totally-symmetric Lorentz indices, $(\mu\nu)$ and $(\mu\nu\rho)$, respectively. 
Note that the metric tensor signature in this work is $g_{\mu\nu}={\rm diag}(1,-1,-1,-1)$.

With using of the standard QFT method \cite{Peskin:1995ev}, the vertices of the interaction Lagrangian in Eqs.(\ref{eqn:pom-lag}) and (\ref{eqn:odd-lag}) are formulated as

\begin{figure}[!htbp]
\centering
\includegraphics[scale=0.5]{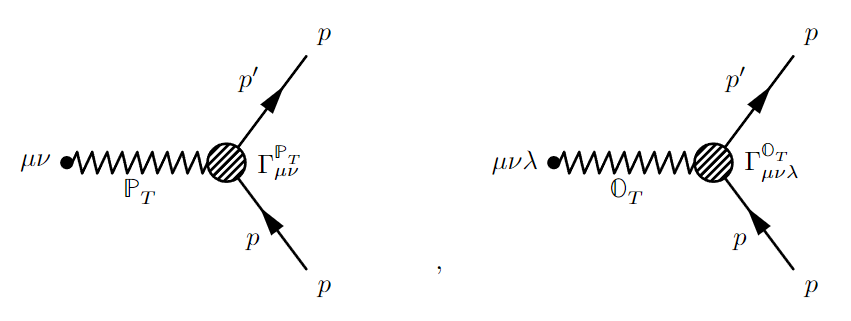}
\vspace{-25pt}
\end{figure}

\begin{align} 
    i\Gamma_{\mu\nu}^{\mathbb{P}_T^{}}(p',p) &= - i \, g_{\mathbb{P}_T^{}}^{} \, \mathcal{G}_{(\mu\nu)}^{\mu'\nu'} \, \gamma_{\mu'}^{} \, (p'+p)_{\nu'}^{} \, \mathcal{F}_{\mathbb{P}_T^{}}^{}\left[(p'-p)^2\right], \label{eqn:pom-ver} \\
    i\Gamma_{\mu\nu\rho}^{\mathbb{O}_T^{}}(p',p) &= - i \, \frac{g_{\mathbb{O}_T^{}}^{}}{{M}_0^{}} \, \mathcal{G}_{(\mu\nu\rho)}^{\mu'\nu'\rho'} \, \gamma_{\mu'}^{} \, (p'+p)_{\nu'}^{} \, (p'+p)_{\rho'}^{} \, \mathcal{F}_{\mathbb{O}_T^{}}^{}\left[(p'-p)^2\right], \label{eqn:odd-ver}
\end{align}
where $p$ and $p'$ are the momentum of incoming and outgoing particles respectively. Also, the coupling constants $g_{\mathbb{P}_T^{}}^{} = 3\beta_{\mathbb{P}NN}^{} = 3 \times 1.87$ GeV$^{-1}$ and $g_{\mathbb{O}_T^{}}^{} = 13.80$ GeV$^{-1}$ are obtained from \cite{Ewerz:2013kda} and \cite{Magallanes2022-odderon}, respectively. The pomeron and odderon form factors are given by
\begin{align}
    \mathcal{F}_{\mathbb{P}_T^{}}^{}(t) &= \left(1-\frac{t}{4\,m_p^2} \frac{\mu_p}{\mu_N}\right) \left( 1 - \frac{t}{4\,m_p^2}\right)^{-1} \left( 1 - \frac{t}{m_D^2}\right)^{-2}, \label{eqn:pom-ff}\\
    \mathcal{F}_{\mathbb{O}_T^{}}^{}(t) &= \left(1-\frac{A\,t}{4\,m_p^2}\frac{\mu_p}{\mu_N}\right)\left( 1 - \frac{B\,t}{4\,m_p^2}\right)^{-1}\left( 1 - \frac{C\,t}{m_D^2}\right)^{-2}, \label{eqn:odd-ff}
\end{align}
with their corresponding normalization of $\mathcal{F}_{\mathbb{P}_T^{}}^{}(0) = 1$ and $\mathcal{F}_{\mathbb{O}_T^{}}^{}(0) = 1$, where the proton mass $m_p^{} = 0.978$ GeV, the ratio of magnetic moment of proton to nucleon $\mu_p^{}/\mu_N^{} = 2.7928$, and the squared dipole mass $m_D^2 = 0.71$ GeV$^2$ are given from \cite{Ewerz:2013kda}. Note that the form factor of the pomeron has no free parameters whereas for odderon the form factor is modified by three parameters: $A = -1.671$, $B = -1.855$, and $C = 0.493$ which is obtained from fitting the model to several experiments, as shown in \cite{Magallanes2022-odderon}.

The corresponding propagators read
\begin{figure}[!htbp]
    \centering
\includegraphics[scale=0.55]{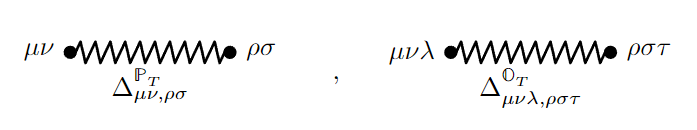}
\vspace{-20pt}
\end{figure}

\begin{align} 
    i\Delta_{\mu\nu,\rho\sigma}^{\mathbb{P}_T^{}}(s,t) &= \frac{1}{4\,s} \left[ g_{\mu\rho}^{} \, g_{\nu\sigma} + g_{\mu\sigma}^{} \, g_{\nu\rho} - \frac{1}{2} g_{\mu\nu}^{} \, g_{\rho\sigma}^{} \right] (-i\, s\, \alpha'_{\mathbb{P}_T^{}})^{\alpha_{\mathbb{P}_T^{}}^{}(t)-1}, \label{eqn:pom-pro}\\
    i\Delta_{\mu\nu\lambda,\rho\sigma\tau}^{\mathbb{O}_T^{}}(s,t) &= -i \, \frac{{M}_0^2}{6 \, s^2}  \left[ \sum_C g_{\mu\rho}^{} (g_{\mu\sigma}^{} \, g_{\lambda\tau}^{} + g_{\nu\tau} \, g_{\lambda\sigma}) - \frac{1}{2} \sum_C g_{\mu\nu}^{} \, g_{\lambda\rho}^{} \, g_{\sigma\tau}^{} \right] (-i\, s\, \alpha'
    _{\mathbb{O}_T^{}})^{\alpha_{\mathbb{O}_T^{}}^{}(t)-1}, \label{eqn:odd-pro}
\end{align}
where $\epsilon$ and $\alpha'$ denote the interception deviation and trajectory slope, respectively. 
For spin-2 tensor pomeron, the trajectory is given by $\alpha_{\mathbb{P}_T^{}}^{}(t) = 1 + \epsilon_{\mathbb{P}_T^{}}^{} + \alpha_{\mathbb{P}_T^{}}' t $ with the $\epsilon_{\mathbb{P}_T^{}}^{} = 0.0808$ and $\alpha_{\mathbb{P}_T^{}}' = 0.25$ GeV$^2$ \cite{Ewerz:2013kda}. 
For spin-3 tensor odderon, the trajectory is given by $\alpha_{\mathbb{O}_T^{}}^{}(t) = 1 + \epsilon_{\mathbb{O}_T^{}}^{} + \alpha_{\mathbb{O}_T^{}}' t $ with the $\epsilon_{\mathbb{O}_T^{}}^{} = 0.062$ and $\alpha_{\mathbb{O}_T^{}}' = 0.189$ GeV$^2$ \cite{Magallanes2022-odderon}.

Next, we will calculate the corresponding helicity amplitudes in the following forms, 
\begin{align} \label{eqn:pom-odd-hel-amps-form}
\begin{aligned}
    i\langle h_3^{} h_4^{} |\mathcal{T}_{\mathbb{P}_T^{}}^{}| h_1^{} h_2^{} \rangle 
        & = \langle p(p_3^{},h_3^{}) p(p_4^{},h_4^{}) |: \mathbb{T} \exp \left( i\int \mathcal{L}_{\mathbb{P}_T^{}}(x) d^4x \right) :| p(p_1^{},h_1^{}) p(p_2^{},h_2^{}) \rangle \\
        & = \bar{u}(p_3^{},h_3^{}) \, i\Gamma_{\mu\nu}^{\mathbb{P}_T^{}}(p_3^{},p_1^{}) \, u(p_1^{},h_1^{}) \, i\Delta_{\mathbb{P}_T^{}}^{\mu\nu,\rho\sigma} (s,t) \, \bar{u}(p_4^{},h_4^{}) \, i\Gamma_{\rho\sigma}^{\mathbb{P}_T^{}}(p_4^{},p_2^{}) \, u(p_2^{},h_2^{}),
\end{aligned} \\
\begin{aligned}
    i\langle h_3^{} h_4^{} |\mathcal{T}_{\mathbb{O}_T^{}}^{}| h_1^{} h_2^{} \rangle 
        & = \langle p(p_3^{},h_3^{}) p(p_4^{},h_4^{}) |: \mathbb{T} \exp \left( i\int \mathcal{L}_{\mathbb{O}_T^{}}(x) d^4x \right) :| p(p_1^{},h_1^{}) p(p_2^{},h_2^{}) \rangle \\
        & = \bar{u}(p_3^{},h_3^{}) \, i\Gamma_{\mu\nu\lambda}^{\mathbb{O}_T^{}}(p_3^{},p_1^{}) \, u(p_1^{},h_1^{}) \, i\Delta_{\mathbb{O}_T^{}}^{\mu\nu\lambda,\rho\sigma\tau} (s,t) \, \bar{u}(p_4^{},h_4^{}) \, i\Gamma_{\rho\sigma\tau}^{\mathbb{O}_T^{}}(p_4^{},p_2^{}) \, u(p_2^{},h_2^{}),
\end{aligned}
\end{align}
where $u(p,h)$ is the Dirac helicity spinor defined in the Appendix \ref{sec:hel-spinor}, $\mathbb{T}$ is the time ordered operator, and $:\;:$ is the normal order operator in the standard QFT \cite{Peskin:1995ev}. 

According to the invariant under the parity conservation ($P$), charge conjugation ($C$), time reversal ($T$), and conservation of total spin, total helicity amplitudes of the two-body fermion-fermion scattering are reduced from sixteen down to five independent amplitudes, as seen in \cite{Donnachie2002-pomeron}. Thus, the scattering amplitudes of the odderon are included, with a minus sign, in the exchange as shown in \cite{Magallanes2022-odderon}. The scattering amplitudes of the $pp$ and $p\bar p$ are written by   
\begin{eqnarray}
\langle h_3^{} h_4^{} |\mathcal{T}_{pp}^{}| h_1^{} h_2^{} \rangle &=& \langle h_3^{} h_4^{} |\mathcal{T}_{\mathbb{P}}^{} - \mathcal{T}_{\mathbb{O}}^{}| h_1^{} h_2^{} \rangle\,,
\\
\langle h_3^{} h_4^{} |\mathcal{T}_{p\bar p}^{}| h_1^{} h_2^{} \rangle &=& \langle h_3^{} h_4^{} |\mathcal{T}_{\mathbb{P}}^{} + \mathcal{T}_{\mathbb{O}}^{}| h_1^{} h_2^{} \rangle\,,
\end{eqnarray}
where $\mathcal{T}_{\mathbb{P}}^{}$ and $\mathcal{T}_{\mathbb{O}}^{}$ are the transition matrices of the pomeron and odderon, respectively.
Consequently, the generic form of the independent helicity amplitudes defined in \cite{Goldberger:1960,Buttimore:1998rj,Donnachie2002-pomeron} as $\phi_i^{} = \langle h_3h_4|\mathcal{T}|h_1 h_2\rangle$ are then decomposed into pomeron and odderon contributions related to the transition matrix $\mathcal{T}$ as given by,
\begin{align} \label{eqn:ind-hel-amps-form}
\begin{aligned}
    \phi_1^{TT}(s,t) = \langle ++ |\mathcal{T}_{pp}^{TT}| ++ \rangle = \langle ++|\mathcal{T}_{\mathbb{P}_T^{}}^{} - \mathcal{T}_{\mathbb{O}_T^{}}^{}|++  \rangle, \\
    \phi_2^{TT}(s,t) = \langle ++ |\mathcal{T}_{pp}^{TT}| -- \rangle = \langle ++|\mathcal{T}_{\mathbb{P}_T^{}}^{} - \mathcal{T}_{\mathbb{O}_T^{}}^{}|-- \rangle, \\
    \phi_3^{TT}(s,t) = \langle +- |\mathcal{T}_{pp}^{TT}| +- \rangle = \langle +-|\mathcal{T}_{\mathbb{P}_T^{}}^{} - \mathcal{T}_{\mathbb{O}_T^{}}^{}|+- \rangle, \\
    \phi_4^{TT}(s,t) = \langle +- |\mathcal{T}_{pp}^{TT}| -+ \rangle = \langle +-|\mathcal{T}_{\mathbb{P}_T^{}}^{} - \mathcal{T}_{\mathbb{O}_T^{}}^{}|-+ \rangle, \\
    \phi_5^{TT}(s,t) = \langle ++ |\mathcal{T}_{pp}^{TT}| +- \rangle = \langle ++|\mathcal{T}_{\mathbb{P}_T^{}}^{} - \mathcal{T}_{\mathbb{O}_T^{}}^{}|+- \rangle,
\end{aligned}
\end{align}
where plus ($+$) and minus ($-$) signs in the states correspond to right-$(\chi_\uparrow)$ and left-$(\chi_\downarrow)$ handed helicities, mentioned in the Appendix \ref{sec:hel-spinor}, respectively. Physically, $\phi_1^{}$ and $\phi_3^{}$ are spin-flip amplitudes, $\phi_2^{}$ and $\phi_4^{}$ are double-flip amplitudes, and $\phi_5^{}$ is single-flip amplitude. The superscript ``$TT$" denotes the consecutive letters as tensors for spin-2 tensor pomeron and spin-3 tensor odderon. Note that we will apply the Regge limit $(s\gg t,m)$ to each amplitude.

Finally, the general form of the ratio of helicity's spin-flip to non-flip amplitudes given by \cite{Buttimore:1998rj}, see also \cite{Ewerz:2016onn}, as
\begin{align} \label{eqn:r5-form}
    r_5^{TT}(s,t) &= \frac{2\,m_p^{} \, \phi_5^{TT}(s,t)} {\sqrt{-t} \, \textrm{Im}[\phi_1^{TT}(s,t) + \phi_3^{TT}(s,t)]}.
\end{align}

In addition, to get a convenient form for presenting the results of this work and comparing them to others, we defined the reduced amplitudes as
\begin{align} \label{eqn:red-hel-amps-form}
\begin{aligned}
    \hat{\phi}_i^{\mathbb{P}_T^{}}(s,t) &= \phi_i^{\mathbb{P}_T^{}}(s,t) / \mathcal{C}_{\mathbb{P}_T^{}}^{}(s,t), \\
    \hat{\phi}_i^{\mathbb{O}_T^{}}(s,t) &= \phi_i^{\mathbb{O}_T^{}}(s,t) / \mathcal{C}_{\mathbb{O}_T^{}}^{}(s,t),
\end{aligned}
\end{align}
where $i = 1,..,5$ is the index of independent helicity amplitudes as used in (\ref{eqn:ind-hel-amps-form}), and the common factors are defined as
\begin{align} \label{eqn:com-factors}
\begin{aligned}
    \mathcal{C}_{\mathbb{P}_T^{}}^{}(s,t) &= i \, g_{\mathbb{P}_T^{}}^2 \mathcal{F}_{\mathbb{P}_T^{}}^2(t)  \frac{1}{4\,s} (-i\,s\,\alpha'_{\mathbb{P}_T^{}})^{\alpha_{\mathbb{P}_T^{}}^{}(t)-1}, \\
    \mathcal{C}_{\mathbb{O}_T^{}}^{}(s,t) &= \frac{8}{9}\, g_{\mathbb{O}_T^{}}^2 \mathcal{F}_{\mathbb{O}_T^{}}^2(t) \frac{1}{6\,s}(-i\,s\,\alpha'_{\mathbb{O}_T^{}})^{\alpha_{\mathbb{O}_T^{}}^{}(t)-1}.
\end{aligned}
\end{align}

For the odderon as a spin-1 vector particle, we, therefore, constructed the \emph{combined spin-1 vector odderon} model in order to calculate and compare the results to that of the \emph{combined spin-3 tensor odderon}.

The Lagrangian of the spin-1 vector odderon from \cite{Ewerz:2013kda} reads
\begin{align} \label{eqn:odd-1-lag}
    \mathcal{L}_{\mathbb{O}_V^{}}^{}(x) &= -  g_{\mathbb{O}_V^{}}^{} M_0^{} \, \bar{\psi}_p(x) \, \gamma_{\mu}^{} \psi_p(x) \, \mathbb{O}^{\mu}(x),
\end{align}
where $g_{\mathbb{O}_V^{}}^{} = $ is the coupling constant, $\mathbb{O}^{\mu}(x)$ is the description of the spin-1 vector odderon. The corresponding vertex reads
\begin{figure}[!htbp]
    \centering
\includegraphics[scale=0.55]{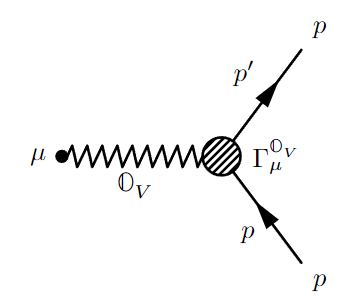}
\vspace{-25pt}
\end{figure}

\begin{align} \label{eqn:odd-1-ver}
    i\Gamma_{\mu}^{\mathbb{O}_V^{}}(p',p) &= -i \, g_{\mathbb{O}_V^{}}^{} \, M_0^{} \, \gamma_{\mu}^{} \, \mathcal{F}_{\mathbb{O}_V^{}}^{}\left[(p'-p)^2\right],
\end{align}
where the form factor is defined the same as the pomeron,
\begin{align} \label{eqn:odd-1-ff}
    \mathcal{F}_{\mathbb{O}_V^{}}(t) &= \left(1-\frac{t}{4m_p^2} \frac{\mu_p}{\mu_N}\right) \left( 1 - \frac{t}{4m_p^2}\right)^{-1} \left( 1 - \frac{t}{m_D^2}\right)^{-2},
\end{align}
with normalization $\mathcal{F}_{\mathbb{O}_V^{}}^{}(0) = 1$. The corresponding propagator reads
\begin{figure}[!htbp]
    \centering
\includegraphics[scale=0.60]{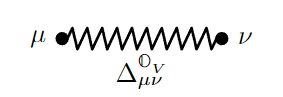}
\vspace{-20pt}
\end{figure}

\begin{align} \label{eqn:odd-1-pro}
    i\Delta_{\mu\nu}^{\mathbb{O}_V^{}}(s,t) &= -i \frac{g_{\mu\nu}^{}}{M_0^2}   (-i\, s\, \alpha'_{\mathbb{O}_V^{}})
    ^{\alpha_{\mathbb{O}_V^{}}^{}(t)-1},
\end{align}
where the odderon trajectory is $\alpha_{\mathbb{O}_V^{}}^{}(t) = 1 + \epsilon_{\mathbb{O}_V^{}}^{} + \alpha'
_{\mathbb{O}_V^{}}t$ with both $\epsilon_{\mathbb{O}_V^{}}^{} = 0.0808$ and $\alpha'_{\mathbb{O}_V^{}} = 0.25$ GeV$^2$ are also taken from \cite{Ewerz:2013kda}. Thus, a form of the corresponding helicity amplitudes are
\begin{align} \label{eqn:odd-1-hel-amps-form}
\begin{aligned}
    i\langle h_3^{} h_4^{} |\mathcal{T}_{\mathbb{O}_V^{}}^{}| h_1^{} h_2^{} \rangle 
        & = \langle p(p_3^{},h_3^{}) p(p_4^{},h_4^{}) |: \mathbb{T} \exp \left( i\int \mathcal{L}_{\mathbb{O}_V^{}}(x) d^4x \right) :| p(p_1^{},h_1^{}) p(p_2^{},h_2^{}) \rangle \\
        & = \bar{u}(p_3^{},h_3^{}) \, i\Gamma_{\mu}^{\mathbb{O}_V^{}}(p_3^{},p_1^{}) \, u(p_1^{},h_1^{}) \, i\Delta_{\mathbb{O}_V^{}}^{\mu\nu} (s,t) \, \bar{u}(p_4^{},h_4^{}) \, i\Gamma_{\nu}^{\mathbb{O}_V^{}}(p_4^{},p_2^{}) \, u(p_2^{},h_2^{}).
\end{aligned}
\end{align}
Consequently, we defined the independent helicity amplitudes of the spin-1 vector odderon in the same manner as the previous case,
\begin{align} \label{eqn:ind-hel-amps-form-1}
\begin{aligned}
    \phi_1^{TV}(s,t) =  \langle ++|\mathcal{T}_{pp}^{TV}|++  \rangle = \langle ++|\mathcal{T}_{\mathbb{P}_T^{}}^{} - \mathcal{T}_{\mathbb{O}_V^{}}^{}|++  \rangle, 
    \\
    \phi_2^{TV}(s,t) = \langle ++|\mathcal{T}_{pp}^{TV}|-- \rangle = \langle ++|\mathcal{T}_{\mathbb{P}_T^{}}^{} - \mathcal{T}_{\mathbb{O}_V^{}}^{}|-- \rangle, 
    \\
    \phi_3^{TV}(s,t)  = \langle +-|\mathcal{T}_{pp}^{TV}|+- \rangle = \langle +-|\mathcal{T}_{\mathbb{P}_T^{}}^{} - \mathcal{T}_{\mathbb{O}_V^{}}^{}|+- \rangle, 
    \\
    \phi_4^{TV}(s,t) = \langle +-|\mathcal{T}_{pp}^{TV}|-+ \rangle 
    = \langle +-|\mathcal{T}_{\mathbb{P}_T^{}}^{} - \mathcal{T}_{\mathbb{O}_V^{}}^{}|-+ \rangle, 
    \\
    \phi_5^{TV}(s,t) = \langle ++|\mathcal{T}_{pp}^{TV}|+- \rangle 
    = \langle ++|\mathcal{T}_{\mathbb{P}_T^{}}^{} - \mathcal{T}_{\mathbb{O}_V^{}}^{}|+- \rangle,
\end{aligned}
\end{align}
where the superscript ``TV" represents the spin-2 tensor pomeron and the spin-1 vector odderon exchanges in $pp$ elastic scattering and the corresponding result of the $r_5$ parameter reads
\begin{align} \label{eqn:r5-form-1}
    r_5^{TV}(s,t) &= \frac{2\,m_p^{} \, \phi_5^{TV}(s,t)} {\sqrt{-t} \, \textrm{Im}[\phi_1^{TV}(s,t) + \phi_3^{TV}(s,t)]}.
\end{align} 
In the same manner, we defined the helicity amplitudes in the following form,
\begin{align} \label{eqn:red-hel-amps-form-1}
    \hat{\phi}_i^{\mathbb{O^{}}_V}(s,t) &= \phi_i^{\mathbb{O}_V^{}}(s,t) / \mathcal{C}_{\mathbb{O}_V^{}}^{}(s,t),
\end{align}
where the corresponding common factor reads
\begin{align} \label{eqn:com-factors-1}
    \mathcal{C}_{\mathbb{O}_V^{}}^{}(s,t) &= g_{\mathbb{O}_V^{}}^2 \, \mathcal{F}_{\mathbb{O}_V^{}}^2(t) \frac{1}{4\,s} (-i\,s\,\alpha'_{\mathbb{O}_V^{}})^{\alpha_{\mathbb{O}_V^{}}^{}(t)-1}.
\end{align}


In this section, we have formulated the relevant observable from the helicity amplitudes in $pp$ elastic scattering. We have also categorized the complex parameter $r_5$ as the results of two models: (1) the $r_5^{TT}$ from the spin-2 tensor pomeron with the spin-3 tensor odderon exchanges and (2) the  $r_5^{TV}$ from the spin-2 tensor pomeron with the spin-1 vector odderon exchanges. We will compare the theoretical results from the $r_5^{TT}$ and $r_5^{TV}$ with the experimental data in the next section.

\section{\label{sec:result} Results and Discussion}

The purpose of this research is to explore the possibility that the odderon with the spin $J=3$, alongside the pomeron, contributes to $pp$ elastic scattering. Therefore, the observable as a complex variable $r_5^{}$, namely the ratio of spin-flip to non-flip amplitudes, of the combined exchange of the odderon and pomeron is evaluated from their corresponding helicity amplitudes.

First, we have calculated the helicity amplitudes of the spin-2 tensor pomeron ($\mathbb{P}_T^{}$) and spin-3 tensor odderon ($\mathbb{O}_T^{}$) contributions, by using (\ref{eqn:ind-hel-amps-form}), in the Regge limit $(s\gg t,m)$. The results are shown in the Table \ref{tab:rphi-odderon}.

\begin{table}[!htbp]
\begin{tabular}{@{}lllllll@{}}
\toprule
& \qquad & $\mathbb{P}_T^{}$ & \qquad  & $\mathbb{O}_T^{}$ & \qquad & $\mathbb{O}_V^{}$ \\
\hline
$\hat{\phi}_1^{}(s,t)$ & \qquad & $8s^2$ & \qquad & $9s^2$ & \qquad & $8s^2$ \\
$\hat{\phi}_2^{}(s,t)$ & \qquad & $10m_p^2t$ & \qquad & $10m_p^2t$ & \qquad & $16m_p^2t$ \\
$\hat{\phi}_3^{}(s,t)$ & \qquad & $8s^2$ & \qquad & $9s^2$ & \qquad & $8s^2$ \\
$\hat{\phi}_4^{}(s,t)$ & \qquad & $-10m_p^2t$ & \qquad & $-10m_p^2t$ & \qquad & $-16m_p^2t$ \\
$\hat{\phi}_5^{}(s,t)$ & \qquad & $-8m_p^{}s\sqrt{-t}$ & \qquad & $-9m_p^{}s\sqrt{-t}$ & \qquad & $-8m_p^{}s\sqrt{-t}$ \\
\botrule
\end{tabular}
\caption{The reduced helicity amplitudes of the pomeron ($\mathbb{P}_T^{}$) and odderon ($\mathbb{O}_T^{}$ and $\mathbb{O}_V^{}$) contributions.}\label{tab:rphi-odderon}%
\end{table}

According to (\ref{eqn:r5-form}), the result of the spin-3 tensor odderon is written as
\begin{align} \label{eqn:r5-result}
\begin{aligned}
    r_5^{TT}(s,t) &= \frac{m_p^2(9\,\mathcal{C}_{\mathbb{O}_T^{}}^{}(s,t) - 8\,\mathcal{C}_{\mathbb{P}_T^{}}^{}(s,t))}{s \, {\rm Im}(8\,\mathcal{C}_{\mathbb{P}_T^{}}^{}(s,t) - 9\,\mathcal{C}_{\mathbb{O}_T^{}}^{}(s,t))} \\
        &= (5.3 - i\,2.2) \times 10^{-5},
\end{aligned}
\end{align}
where the numerical result in the last line comes from the STAR experiment at RHIC \cite{STAR:2012fiw} as $\sqrt{s} = 200$ GeV and no $t$-dependent ($t=0$).

Similarly we have worked out the spin-1 vector odderon ($\mathbb{O}_V$) contribution. The helicity amplitudes from (\ref{eqn:ind-hel-amps-form-1}) are calculated using the results in the Table \ref{tab:rphi-odderon}.
Thus, the corresponding result according to (\ref{eqn:r5-form-1}) is given by
\begin{align} \label{eqn:r5-o1-result}
\begin{aligned}
        r_5^{TV}(s,t) &= \frac{m_p^2(\mathcal{C}_{\mathbb{O}_V^{}}^{}(s,t) - \mathcal{C}_{\mathbb{P}_T^{}}^{}(s,t))}{s \, {\rm Im}(\mathcal{C}_{\mathbb{P}_T^{}}^{}(s,t) - \mathcal{C}_{\mathbb{O}_V^{}}^{}(s,t))}, \\
        &= (1.7 - i\,2.2) \times 10^{-5}.
\end{aligned}
\end{align}

\begin{figure}[!htbp]
\includegraphics[width=0.7\textwidth]{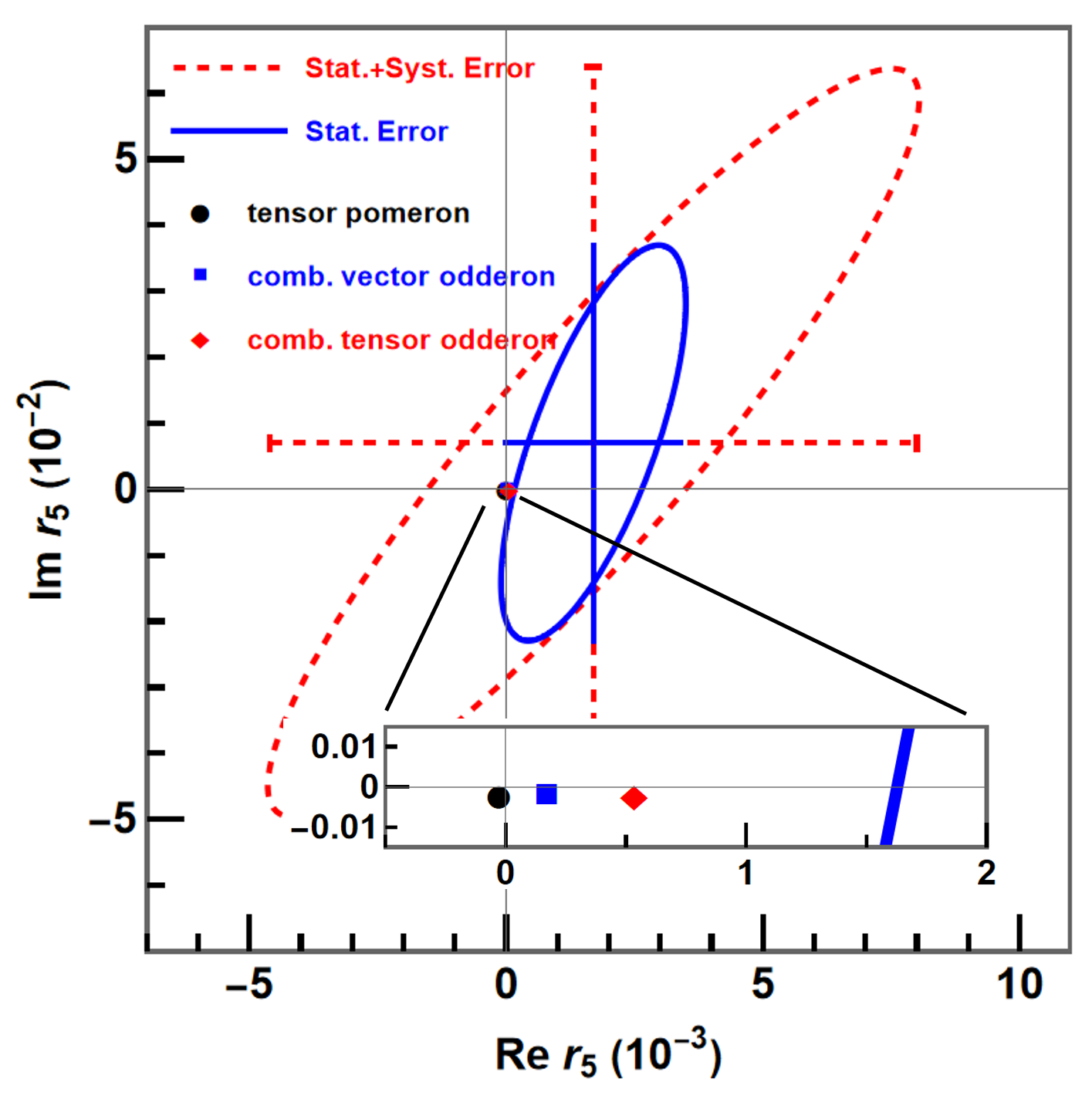}
\caption{\label{fig:out-r5-pom}  (color online) The statistical (solid ellipse and cross) and statistical+systematic uncertainties (dashed ellipse and cross) of the $r_5^{}$ \cite{STAR:2012fiw}. The \emph{pure spin-2 tensor pomeron} (black circle) \cite{Ewerz:2013kda} is $r_5^{\mathbb{P}_T^{}}$. The \emph{combined spin-1 vector odderon} (blue square) is $r_5^{TV}$ and \emph{combined spin-3 tensor odderon} (red diamond) is $r_5^{TT}$.}
\end{figure}

We then evaluated our results of $r_5^{TT}$ by comparing to that of the spin-2 tensor pomeron studied in \cite{Ewerz:2016onn}. The constraints on the $r_5$ from the STAR experiment at RHIC \cite{STAR:2012fiw} are applied to these results for their validity. As shown in the Fig. \ref{fig:out-r5-pom}, the central value of the complex parameter $r_5^{}$ is obtained from the experimental study of spin-single asymmetry $A_N^{}$ of polarized $pp$ elastic scattering at $\sqrt{s} = 200$ GeV \cite{STAR:2012fiw} as $(170.0 + i\,170.0) \times 10^{-5}$ with the corresponding statistical uncertainties $\sigma_{\rm stat} = \pm (170.0 + i\,300.0) \times 10^{-5}$ and statistical+systematic uncertainties $\sigma_{\rm stat + sys} = \pm (630.0 + i\,570.0) \times 10^{-5}$. 

The value from the \emph{pure spin-2 tensor pomeron} model is given as a benchmark, $r_5^{\mathbb{P}_T^{}} = -(0.28 + i\,2.2) \times 10^{-5}$ \cite{Ewerz:2016onn}. Notice that the real part of $r_5^{\mathbb{P}_T^{}}$ is negative which is similar to the result from the study of QCD-inspired dipole model \cite{Hagiwara2020:80-427}. The value lies outside the range of the statistical uncertainties while it remains inside the statistical+systematic uncertainties. 

The main result of this work is the value of $r_5^{TT}$ of the \emph{combined spin-3 tensor odderon} model which has a positive real part. This comes from the fact that the contribution of the odderon has the opposite sign against the pomeron, as defined in (\ref{eqn:ind-hel-amps-form}) and (\ref{eqn:ind-hel-amps-form-1}), see also \cite{Donnachie2002-pomeron,Magallanes2022-odderon}. This value is compatible with the constraints given from the STAR experiment at RHIC \cite{STAR:2012fiw}. Importantly, the clear tendency toward the central value obtained from the experiment indicates the improvement of the model compared to the \emph{pure spin-2 tensor pomeron} model \cite{Ewerz:2016onn}.


In addition, the value of $r_5^{TV}$ of the \emph{combined spin-1 vector odderon} model is properly valid the same order as in the \emph{pure spin-2 tensor pomeron} and \emph{combined spin-3 tensor odderon} models. Superficially, the real part has the positive value as the previous case. However, despite the tendency, its distance to the central value obtained from the experiment is farther than the result of the \emph{combined spin-3 tensor odderon} model. This means that the spin of odderon is likely to be $J=3$ rather than $J=1$.

\section{Conclusion}\label{sec:conclusion}
In spite of the discovery, the spin-dependence aspect of the odderon contribution in $pp$ elastic scattering has not yet been thoroughly investigated. Due to the fact that the odderon is respected as a charge-conjugation partner of pomeron, the observable of the model naturally requires the inclusion of both pomeron and odderon contributions.

Therefore, we have constructed the combined models of spin-3 tensor odderon and spin-1 vector odderon with the spin-2 tensor pomeron by using the DL ansatz for establishing the propagator forms. In each model, the complex parameter $r_5^{}$, as a result, has been calculated from the helicity amplitudes of the models. Both models have been applied to the constraints given from the STAR experiment at RHIC \cite{STAR:2012fiw}. In detail, at first, we have reproduced the parameter $r_5^{\mathbb{P}_T^{}}$ of the \emph{pure spin-2 tensor pomeron} with the free parameters given from \cite{Ewerz:2016onn}. The result of the spin-2 tensor pomeron is identical to that of in \cite{Ewerz:2016onn} and compatible with the study of \cite{Hagiwara2020:80-427}. Next, the parameter $r_5^{TT}$ have also been calculated from the combined helicity amplitudes of the spin-2 tensor pomeron and the spin-3 tensor odderon. This \emph{combined spin-3 tensor odderon} model is the main target of our study. The result of the \emph{combined spin-3 tensor odderon} model is valid with the constraints obtained from the experiment. Also, given that the positive sign of the Re $r_5^{TT}$ against the negative sign of the Re $r_5^{\mathbb{P}_T^{}}$ reflects the charge conjugation partner to the contribution of the $C$-even pomeron and $C$-odd odderon. Then, we worked out the parameter $r_5^{TV}$ of the spin-1 vector odderon in the same manner as the spin-3 tensor odderon. We have found that, despite the same tendency, the result of the \emph{combined spin-1 vector odderon} model is farther from the central value than the \emph{combined spin-3 tensor odderon} model. Finally, we plotted all the results with the constraints from the experiment and drawn the conclusion that the spin-3 tensor odderon is more likely to contribute to $pp$ elastic scattering.



{The contribution of the pomeron and odderon are expected to be dominated at TeV energy scale in $pp$ elastic scattering. In fact the parameters of our model have been chosen from the best fit of various experiments \cite{Magallanes2022-odderon}. However, in this work the model is subjected to the constrain from the lower energy experiment of $\sqrt{s} = 200$ GeV which is the only available data for the observable $r_5^{}$ of spin-dependence $pp$ elastic scattering at the present time. Therefore the future experiments at higher energy would shine the light on the odderon contribution in helicity amplitudes}

{The fact that our results are still outside the statistical error of the constraints given from the STAR experiment at RHIC indicates that the other relevant exchanges could be introduced into the reaction at lower energy scale in order to explain the experimental result. Therefore, as described in \cite{Ewerz:2013kda}, the other contributions for the exchange in the reaction, such as $f_{2R}^{}, a_{2R}^{}, \omega_R^{}, \rho_R^{}$ and $\gamma$, are of interest to study further and we leave these topics for future works.}

\appendix
\section{The Helicity Spinors and Kinematics} \label{sec:hel-spinor}
The central results in this work are helicity amplitudes of $pp$ elastic scattering with the pomeron and odderon exchanges. Therefore, in order to make the calculations in this work transparently and well organized, it is worth fixing the normalization of the Dirac spinors and their helicity conventions. With $\bar u_r(p,m)\,u_s(p,m)=2m\delta_{rs}$ normalization, one can write  
the Dirac helicity spinors as
\begin{align}
\begin{aligned}
    u_{\uparrow}^{}(p,m) &= \sqrt{E+m}
    \begin{bmatrix}
        \chi_{\uparrow}^{} \\
        \kappa \chi_{\uparrow}^{}
    \end{bmatrix}, \quad\quad
    u_{\downarrow}^{}(p,m) = \sqrt{E+m}
    \begin{bmatrix}
        \chi_{\downarrow}^{} \\
        -\kappa \chi_{\downarrow}^{}
    \end{bmatrix}, \\
    v_{\uparrow}^{}(p,m) &= \sqrt{E+m}
    \begin{bmatrix}
        -\kappa \chi_{\downarrow}^{} \\
        \chi_{\downarrow}^{}
    \end{bmatrix}, \quad\quad
    v_{\downarrow}^{}(p,m) = \sqrt{E+m}
    \begin{bmatrix}
        \kappa \chi_{\uparrow}^{} \\
        \chi_{\uparrow}^{} 
    \end{bmatrix},
\end{aligned}
\end{align}
where $\kappa = \frac{\lvert\vec{p}\rvert}{E+m} = \frac{\gamma\beta}{\gamma+1}$, $\gamma = 1/\sqrt{1-\beta^2}$, and $\beta=\sqrt{1-4m^2/s}$. While the two-component spinors read
\begin{align}
    \chi_{\uparrow}^{} = 
    \begin{bmatrix}
        \cos(\theta/2) \\
        e^{i\phi}\sin(\theta/2)
    \end{bmatrix}, \quad\quad
    \chi_{\downarrow}^{} = 
    \begin{bmatrix}
        -\sin(\theta/2) \\
        e^{i\phi}\cos(\theta/2)
    \end{bmatrix},
\end{align}
where $\phi$ is the angle in the spin-related plane of the particles.
The four momenta of the particles are read, (Figure \ref{fig:fd-pp-scattering})
\begin{align}
\begin{aligned}
    p_1^{\mu} &= \frac{\sqrt{s}}{2} (1,0,0,\beta), \\
    p_2^{\mu} &= \frac{\sqrt{s}}{2} (1,0,0,-\beta), \\
    p_3^{\mu} &= \frac{\sqrt{s}}{2} (1,\beta\sin{\theta},0,\beta\cos{\theta}), \\
    p_4^{\mu} &= \frac{\sqrt{s}}{2} (1,-\beta\sin{\theta},0,-\beta\cos{\theta}).
\end{aligned}
\end{align}

In addition, all momenta are fixed the directions on the c.m. frame scattering plane as 
\begin{align}
\begin{aligned}
    p_1^{} &: \, \theta=0 \,,\, \phi=0 \,, \\
    p_2^{} &: \, \theta=\pi \,,\, \phi=\pi \,, \\
    p_3^{} &: \, \theta \,,\, \phi=0 \,, \\
    p_4^{} &: \, \theta=\pi-\theta \,,\, \phi=\pi. 
\end{aligned}
\end{align}
In the Regge limit $(s\gg t,m)$, moreover, the trigonometric functions of the c.m. frame angle can be written in terms of the Mandelstam variables as
\begin{align} \label{eqn:trig-kin-relations}
\begin{aligned}
    \sin({\theta/2}) &\approx \sqrt{-t/s}, \\
    \cos({\theta/2}) &\approx 1, \\
    \sin(\theta) &\approx 2\sqrt{-{t/s}}, \\
    \cos(\theta) &\approx 1, \\
    \cos(2\theta) &\approx 0.
\end{aligned}
\end{align}
We will use all notations and conventions introduced above carry out the calculations throughout this work.
\acknowledgments
DS is supported by the Fundamental Fund of Khon Kaen University and DS has received funding support from the National Science, Research and Innovation Fund. The Mindanao State University - Iligan Institute of Technology is also acknowledged through its research and extension support extended to J.B. Magallanes for his travel to Khon Kaen University, Thailand. PS, CP, and DS are financially supported by the National Astronomical Research Institute of Thailand (NARIT). CP and DS are supported by Thailand NSRF via PMU-B [grant number B05F650021]. CP is also supported by Fundamental Fund 2565 of Khon Kaen University and Research Grant for New Scholar, Office of the Permanent Secretary, Ministry of Higher Education, Science, Research and Innovation under contract no. RGNS64-043. 

The authors acknowledge the National Science and Technology Development Agency, National e-Science Infrastructure Consortium, Chulalongkorn University and the Chulalongkorn Academic Advancement into Its 2nd Century Project, NSRF via the Program Management Unit for Human Resources \& Institutional Development, Research and Innovation [grant numbers B05F650021, B37G660013] (Thailand) for providing computing infrastructure that has contributed to the research results reported within this paper. URL:www.e-science.in.th.


\bibliography{bibs}

\end{document}